\documentclass[preprint,aps,prd,showpacs,nofootinbib,superscriptaddress,tightenlines]{revtex4}

\usepackage{amsmath}

\def\slash#1{#1 \hskip-0.45em /}
\def\Slash#1{#1 \hskip-0.59em /}
\newcommand{\nn}{\nonumber}
\allowdisplaybreaks[1]

\begin{document}
\title{The leading twist light-cone distribution amplitudes for the S-wave and P-wave $B_c$ mesons}

\author{Ji Xu\footnote{E-mail: xuji13@mails.ucas.ac.cn}}
\affiliation{School of Physical Sciences, University of
Chinese Academy of Sciences, Yuquan Road 19A, Beijing 100049, China\vspace{0.2cm}}

\author{Deshan Yang\footnote{E-mail: yangds@ucas.ac.cn}}
\affiliation{School of Physical Sciences, University of Chinese Academy of Sciences, Yuquan Road 19A, Beijing 100049, China\vspace{0.2cm}}
\affiliation{Institute of High Energy Physics, 
Chinese Academy of Sciences, Yuquan Road 19B, Beijing 100049, China}

\date{July 19, 2016}

\vspace{1cm}
\begin{abstract} The light-cone distribution amplitudes (LCDAs) serve as important non-perturbative inputs for the study of hard exclusive processes. 
In this paper,  we calculate ten LCDAs at twist-2 for the S-wave and P-wave $B_c$ mesons up to the next-to-leading order (NLO) of the strong coupling $\alpha_s$ and leading order of the velocity expansion. Each one of these ten LCDAs is expressed as a product of a perturbatively calculable distribution and a universal NRQCD matrix-element. By use of the spin symmetry, only two NRQCD matrix-elements will be involved. The reduction of the number of non-perturbative inputs will improve the predictive power of collinear factorization.

\end{abstract}


\maketitle

\section{Introduction}
$B_c$ meson family is unique since they are composed of two different flavors of heavy quark and anti-quark.  
Such uniqueness attracts a lot of attentions from both experimentalists \cite{Abe:1998fb,Abe:1998wi,Abulencia:2006zu,LHCb:2012ag} and theorists \cite{Chang:1992pt,Chang:1996jt,Chang:2001pm,Wu:2002ig}. 

The main difficulty to study the productions and decays of $B_c$ mesons is that there are too many energy scales entangled in these processes.  However, on the other hand, this entanglement of the many scales  provides an excellent platform for testing many fascinating aspects of the perturbative QCD and factorizations. Generally, we think that $B_c$ meson is a non-relativistic bound-state of $b$ quark and $\bar c$ quark, similar to quarkonium. Thus, many researches on $B_c$ involved processes are done within the framework of non-relativistic QCD (NRQCD) factorization (for reviews of NRQCD, see \cite{Bodwin:1994jh, Brambilla:2004jw}), such as some recent studies on $B_c$ decays, $B_c$ productions from $W$ and $Z$ boson decays, etc \cite{Qiao:2011yk,Qiao:2011yz,Qiao:2011zc,Qiao:2012vt,Qiao:2012hp,Jiang:2015jma,Chen:2015csa,Jiang:2015pah}. In NRQCD factorization, the cross-section for inclusive process or the amplitude for exclusive process can be factorized into a product of the perturbatively calculable short-distance coefficient and the non-perturbative NRQCD matrix-element.  
For $B_c$ production at large momentum transfer, say $Q^2>>m_{B_c}^2$,  the short-distance coefficient in NRQCD factorization involves at least two well-separated scales $Q^2$ and $m_{B_c}^2$. This causes large logarithmic terms like $\alpha_s^n\ln^n Q^2/m_{B_c}^2$ in perturbation series, which cannot be resummed within the NRQCD method. This is quite similar to the problems in quarkonia production at large momentum transfer. To cure such problems, one needs to either re-factorize the short-distance coefficients in some way, or adopt another factorization scheme, for instance the collinear factorization, to calculate such productions.  

Taking the exclusive process $\gamma^*(q) B_c^-(p_1)\to B_c^-(p_2)$ at large momentum transfer $Q^2\equiv-q^2=-(p_2-p_1)^2>>m_{B_c}^2$ as an example, 
 the electromagnetic (EM) transition form-factor $F(Q^2)$ parameterizing the amplitude can have a similar collinear factorization formula to the EM transition form-factor for $\gamma^* \pi^-\to \pi^-$ at large momentum transfer \cite{PhysRevD.22(1980),Chernyak:1983ej}, which reads as
\begin{eqnarray}\label{eq:factBc}
F(Q^2)=\int_0^1 dx\, \int_0^1 dy\, \Phi_{B_c}(y;\mu_F)T_H(x,y,Q^2;\mu_R,\mu_F)\Phi_{B_c}(x;\mu_F)+{\cal O}(1/Q^4)\,,
\end{eqnarray}
where $\Phi_{B_c}(x;\mu_F)$ is the light-cone distribution amplitude (LCDA) for $B_c$ meson at twist-2, and $T_H(x,y)$ represents the perturbatively calculable hard-scattering kernel with $x$ and $y$ being the light-cone momentum fractions. The LCDA  $\Phi_{B_c}(x,\mu_F)$ encodes the hadronization effects, and obeys the celebrated Efremov-Radyushkin-Brodsky-Lepage (ERBL) equation \cite{Lepage:1979zb,Efremov:1979qk}
\begin{eqnarray}
 \mu \frac{d}{d \mu} \Phi_{B_c}(x;\mu)=\frac{\alpha_s(\mu)}{\pi}C_F\int_0^1 V_0(x,y)\Phi_{B_c}(y;\mu),
\end{eqnarray}
where $V_0(x,y)$ is the Brodsky-Lepage kernel.

Different from the LCDAs for light mesons which are totally non-perturbative objects, the LCDA for quarkonium or $B_c$ meson can be further factorized  
into a product of a perturbatively calculable distribution part and a NRQCD matrix-element for the vacuum to quarkonium or $B_c$ state transition, at the leading order of non-relativistic expansion parameter $v$  \cite{Ma:2006hc,Bell:2008er}.  As an examination,  it has been shown in \cite{Jia:2010fw} that the factorization formula (\ref{eq:factBc}) by employing $\Phi_{B_c}(x;\mu)$ given in \cite{Bell:2008er}  do reproduce the asymptotic behavior of the EM transition form-factor calculated directly from the NRQCD computation, up to the next-to-leading order (NLO) of the strong coupling $\alpha_s$ and leading order of $v$-expansion. Thus, in this sense, the re-factorization of the LCDAs for $B_c$ meson can serve as a bridge to connect the NRQCD factorization and collinear factorization for $B_c$ meson exclusive productions at large momentum transfer.   

In this paper, we extend the computation of the leading twist LCDA for the S-wave pseudo-scalar $B_c$ meson in \cite{Bell:2008er} to 
all the ten twist-2 LCDAs for S-wave and P-wave $B_c$ mesons, up to the NLO of $\alpha_s$ and leading order of $v$.  This can also be regarded as a succeeding work of \cite{Jia:2010fw,Wang:2013ywc}, in which the LCDAs for quarkonia are calculated in the same manner. 

This paper is organized as follows: in section \ref{sect:Definitions}, we give the notations we use in this paper, and present the definitions of the leading twist LCDAs for the S-wave and P-wave $B_c$ mesons in terms of the matrix-elements of a certain class of non-local QCD operators, then we give tree-level results for these LCDAs at the leading order of $v$ within the NRQCD factorization;  in section \ref{sect:NLO}, we present our results of the LCDAs at the NLO of $\alpha_s$ and leading order of $v$ followed by some useful inverse moments of these LCDAs;  finally, we discuss some potential applications of our results and summarize our work in section \ref{sect:summary}.

\section{The definitions of LCDAs for $B_c$ mesons\label{sect:Definitions}}
\subsection{Notations}
We adopt the same notations as in \cite{Wang:2013ywc}: the momentum of $B_c$ meson is $P^\mu=m_{B_c}v^\mu$ where $v^\mu$ is the velocity four-vector of $B_c$ satisfying $v^2=1$.  A 4-vector $a$ can be decomposed as $a^\mu=v\cdot a v^\mu +a_\top^\mu$ where $v\cdot a_\top\equiv 0$. Note that we also use the same notation $v$ to denote the non-relativistic relative velocity between quark and anti-quark inside a rest $B_c$ meson, which characterizes the velocity expansion in the NRQCD. We remind that the readers should not confuse these two in the context. Two light-like vectors $n_{\pm}$ are also introduced, $n_{\pm}^2=0$ and $n_+n_-=2$. A 4-vector $a$ can be decomposed as $a^\mu= n_+a n_-^\mu/2+n_-a n_+^\mu/2+a_\perp^\mu$ with $n_{\pm}a_\perp\equiv0$. For convenience, we set $v^\mu=(n_+vn_-^\mu+n_-vn_+^\mu)/2$ (apparently $n_+vn_-v=1$).

\subsection{Definitions of the LCDAs}
The leading twist LCDAs for the S-wave and P-wave $B_c$ mesons are defined as the matrix-elements of the proper gauge invariant non-local quark bilinear operators
\begin{eqnarray}
   J[\Gamma](\omega)\equiv (\bar b W_c)(\omega n_+)\slash{n}_+\Gamma(W_c^\dag c)(0)\,,
\end{eqnarray}
where $b$ and $c$ are the fields in QCD for $b$ quark and $c$ quark, respectively. The Wilson-line
\begin{eqnarray}
W_c(x)=\textrm{P  exp} \Big( ig_s\int_{-\infty}^0 ds~ n_+A(x+sn_{+})  \Big)\,,
\end{eqnarray}
is a path-ordered exponential with the path along the $n_+$ direction where $g_s$ is the SU(3) gauge coupling and $A_\mu\equiv A_\mu^a(x)T^a$ ($T^a$ are generators of SU(3) group in the fundamental representation).

The ten non-vanishing twist-2 LCDAs for the S-wave and P-wave $B_c$ mesons are defined as
\begin{subequations}\label{eq:def1}
\begin{eqnarray}
\langle B_c(^1S_0,P)\vert J[\gamma_5](\omega)\vert 0\rangle&=&- i f_{P}n_+P\int_0^1 dx ~e^{i\omega n_+Px} \hat{\phi}_{P}(x;\mu)\,, \\
\langle B_c(^3S_1,P,\varepsilon^*)\vert J[1](\omega)\vert 0\rangle&=&- i f_{V}m_V n_+\varepsilon^*\int_0^1dx~e^{i\omega n_+Px} \hat{\phi}_{V}^{\parallel}(x;\mu)\,,\\
\langle B_c(^3S_1,P,\varepsilon^*)\vert
J[\gamma^\alpha_\perp](\omega)\vert 0\rangle&=& -i f_{V}^{\perp}n_+P
\varepsilon_\perp^{*\alpha} \int_0^1 dx ~e^{i\omega n_+Px}  \hat{\phi}_{V}^{\perp}(x;\mu)\,,  \\
\langle B_c(^1P_1,P,\varepsilon^*)\vert J[\gamma_5](\omega)\vert 0\rangle&=& i f_{1A}m_{1A}n_+\varepsilon^* \int_0^1 dx ~e^{i\omega n_+Px}
\hat{\phi}_{1A}^{\parallel}(x;\mu)\,,\\
\langle B_c(^1P_1,P,\varepsilon^*)\vert
J[\gamma^\alpha_\perp\gamma_5](\omega)\vert 0\rangle&=& i f_{1A}^{\perp}n_+P
\varepsilon_\perp^{*\alpha} \int_0^1 dx ~e^{i\omega n_+Px} \hat{\phi}_{1A}^{\perp}(x;\mu)\,,\\
\langle B_c(^3P_0,P)\vert J[1](\omega)\vert
0\rangle&=&
f_{S} n_+P \int_0^1 dx ~e^{i\omega n_+Px}  \hat{\phi}_{S}(x;\mu)\,,\\
\langle B_c(^3P_1,P,\varepsilon^*)\vert J[\gamma_5](\omega)\vert 0\rangle&=&i f_{3A} m_{3A}n_+\varepsilon^*\int_0^1 dx ~e^{i\omega n_+Px}
\hat{\phi}_{3A}^{\parallel}(x;\mu)\,,\\
\langle B_c(^3P_1,P,\varepsilon^*)\vert
J[\gamma^\alpha_\perp\gamma_5](\omega)\vert 0\rangle&=& i f_{3A}^\perp n_+P
\varepsilon_\perp^{*\alpha} \int_0^1 dx ~e^{i\omega n_+Px}  \hat{\phi}_{3A}^{\perp}(x;\mu)\,,\\
\langle B_c(^3P_2,P,\varepsilon^*)\vert J[1](\omega)\vert
0\rangle&=&
f_{T} \frac{m_T^2 }{n_+P}n_{+\alpha}n_{+\beta}\varepsilon^{*\alpha\beta} \int_0^1 dx ~e^{i\omega n_+Px}   \hat{\phi}_{T}^{\parallel}(x;\mu)\,,\\
\langle B_c(^3P_2,P,\varepsilon^*)\vert
J[\gamma^\alpha_\perp](\omega)\vert 0\rangle&=& f_{T}^{\perp} m_T n_{+\rho}
\varepsilon^{*\rho\alpha_\perp} \int_0^1 dx ~e^{i\omega n_+Px}  \hat{\phi}_{T}^{\perp}(x;\mu)\,.
\end{eqnarray}
\end{subequations}
Here $f, \varepsilon^*$ and $\hat \phi(x)$ are decay constants, polarization vectors/tensors, and the distribution parts of twist-2 LCDAs for corresponding $B_c$ mesons, respectively. $x\in[0,1]$ denotes the light-cone fraction (we will also adopt the notation $\bar x\equiv 1-x$ in the rest of this paper),  and $\mu$ is the renormalization scale .

We set the normalization conditions for the LCDAs as following
\begin{subequations}\label{condition}
\begin{eqnarray}
&&\int_0^1 dx \hat\phi_{P} (x)=\int_0^1 dx \hat\phi_{V}^\parallel (x)=\int_0^1 dx \hat\phi_{V}^\perp(x)=\int_0^1 dx \hat\phi_{1A}^\perp (x)=\int_0^1 dx \hat\phi_{3A}^\parallel (x)=1\,,\label{condition1}
\\
  &&\int_0^1 dx  \hat\phi_{1A}^{\parallel}(x)=\int_0^1 dx \hat\phi_{3A}^{\perp}(x)=\int_0^1 dx \hat\phi_{S}(x)=1     \,,\label{condition2}\\
  &&\int_0^1 dx \hat\phi_{T}^{\parallel}(x)(2x-1)=\int_0^1 dx \hat\phi_{T}^{\perp}(x)(2x-1)=1 \,.       \label{condition3}
\end{eqnarray}
\end{subequations}
Note that the normalization conditions for $\hat{\phi}^{\parallel}_{1A}(x)$, $\hat{\phi}^\perp_{3A}(x)$ and $\hat{\phi}_S(x)$ in (\ref{condition2}) are different from the ones for quarkonia \cite{Wang:2013ywc}.  This is due to the fact that we have no C-parity constraints on the LCDAs for $B_c$ mesons.

In the calculation below, we actually use the Fourier transformed forms of the definitions given in (\ref{eq:def1}) such that
\begin{subequations}\label{eq:def2}
\begin{eqnarray}
  \langle B_c(^1S_0,P)\vert Q[\gamma_5](x)\vert 0\rangle&=&- i f_{P} \hat{\phi}_{P}(x)\,,\\
\langle B_c(^3S_1,P,\varepsilon^*)\vert Q[1](x)\vert 0\rangle&=&- i f_{V} \frac{n_+\varepsilon^*}{n_+v}\hat{\phi}_{V}^{\parallel}(x)\,,\\
\langle B_c(^3S_1,P,\varepsilon^*)\vert
Q[\gamma^\alpha_\perp](x)\vert 0\rangle&=& -i f_{V}^{\perp}
\varepsilon_\perp^{*\alpha} \hat{\phi}_{V}^{\perp}(x)\,,  \\
\langle B_c(^1P_1,P,\varepsilon^*)\vert Q[\gamma_5](x)\vert 0\rangle&=&- i f_{1A} \frac{{n_+\varepsilon^*}}{n_+v}
\hat{\phi}_{1A}^{\parallel}(x)\,,\\
\langle B_c(^1P_1,P,\varepsilon^*)\vert
Q[\gamma^\alpha_\perp\gamma_5](x)\vert 0\rangle&=& -i f_{1A}^{\perp}
\varepsilon_\perp^{*\alpha} \hat{\phi}_{1A}^{\perp}(x)\,,\\
\langle B_c(^3P_0,P)\vert Q[1](x)\vert
0\rangle&=&
f_{S}  \hat{\phi}_{S}(x)\,,\\
\langle B_c(^3P_1,P,\varepsilon^*)\vert Q[\gamma_5](x)\vert 0\rangle&=&- i f_{3A} \frac{{n_+\varepsilon^*}}{n_+v}
\hat{\phi}_{3A}^{\parallel}(x)\,,\\
\langle B_c(^3P_1,P,\varepsilon^*)\vert
Q[\gamma^\alpha_\perp\gamma_5](x)\vert 0\rangle&=& -i f_{3A}^\perp
\varepsilon_\perp^{*\alpha} \hat{\phi}_{3A}^{\perp}(x)\,,\\
\langle B_c(^3P_2,P,\varepsilon^*)\vert Q[1](x)\vert
0\rangle&=&
f_{T} \frac{n_{+\alpha}n_{+\beta}\varepsilon^{*\alpha\beta}}{(n_+v)^2}\hat{\phi}_{T}^{\parallel}(x)\,,\\
\langle B_c(^3P_2,P,\varepsilon^*)\vert
Q[\gamma^\alpha_\perp](x)\vert 0\rangle&=& f_{T}^{\perp} \frac{n_{+\rho}
\varepsilon^{*\rho\alpha_\perp}}{n_+v} \hat{\phi}_{T}^{\perp}(x)\,, 
\end{eqnarray}              
\end{subequations}
where 
\begin{eqnarray}
 Q[\Gamma](x)&\equiv&\bigg[      (\bar b W_c)(\omega n_+)\slash{n}_+\Gamma(W_c^\dag c)(0)       \bigg]_{\rm F.T. }    \nn\\
&=&\int\frac{d\omega}{2\pi}e^{-ixn_+P\omega}(\bar b W_c)(\omega n_+)\slash{n}_+\Gamma(W_c^\dag c)(0) .
\end{eqnarray}
Here we suppress the dependence of all quantities on the renormalization scale $\mu$. One can see clearly that all the LCDAs are defined in a boost-invariant way, i.e. they are invariant under the Lorentz boost in which a 4-vector $a^\mu$ transforms as $n_+a\to \alpha n_+a$, $n_-a\to \alpha^{-1}n_-a$ and $a_\perp^\mu\to a_\perp^\mu$.  

\subsection{NRQCD factorization for the LCDAs}

In \cite{Ma:2006hc}, the authors show that the LCDA for $\eta_c$ or $J/\psi$ can be factorized into the product of a perturbative function and a NRQCD matrix-element, since quarkonium is a non-relativistic bound-state of heavy quark and anti-quark.  Taking advantage of the similar nature of $B_c$ meson as a non-relativistic bound-state of $b$ and $\bar c$ quark, we can reduplicate their calculations in the case of $B_c$ mesons. 

Here,  since we are going to deal with two different flavors of heavy quarks, we have to    introduce two kinds of heavy quark effective fields in the NRQCD Lagrangian. By use of the four-component notations as in \cite{Beneke:2008pi},  we have the leading order NRQCD Lagrangian
\begin{eqnarray}\label{eq:NRQCDL}
{\cal L}_{\rm NRQCD}^{\rm LO}=\sum\limits_{Q=b,c}\left[\bar\psi_{Qv}\left(iv\cdot D-\frac{\left(i D^{\mu}_\top\right)\left(i D_{\top\mu} \right)}{2 m_Q}\right)\psi_{Qv}+\bar\chi_{Qv}\left(iv\cdot D+\frac{\left(i D^{\mu}_\top\right)\left(i D_{\top\mu} \right)}{2 m_Q}\right)\chi_{Qv}\right]\,.\nn\\
\end{eqnarray}
Here $m_Q$ is the pole mass of the heavy quark $Q$ ($Q=b,c$), $\psi_{Qv}$ and $\chi_{Qv}$ are the effective fields of the heavy-quark and anti-heavy-quark, respectively, satisfying $\slash v\psi_{Qv}=\psi_{Qv}$ and $\slash v\chi_{Qv}=-\chi_{Qv}$. $D^\mu= \partial^\mu-ig_s A^\mu$ is the covariant derivative.

Schematically, at operator level, we have the matching equation
\begin{eqnarray}
  Q[\Gamma](x,\mu)=\sum\limits_{n=0}^\infty C_\Gamma^n (x,\mu) O_{\Gamma,n}^{\rm NRQCD} \,,
\end{eqnarray}
where $n$ denotes the order of $v$-expansion, $C_\Gamma^n (x,\mu)$ is the short-distance coefficient as a distribution over the light-cone fraction $x$, and $ O_{\Gamma,n}^{\rm NRQCD}$ is the relevant NRQCD operator which scales $\mathcal{O}(v^n)$ in the NRQCD power-counting. So, the LCDAs of $B_c$ meson can be expressed as
\begin{eqnarray}\label{eq:matching}
  \langle B_c\vert Q[\Gamma](x,\mu)\vert 0\rangle\simeq \sum\limits_{n=0}^\infty C_\Gamma^n(x,\mu)\langle B_c\vert O_{\Gamma,n}^{\rm NRQCD} \vert 0\rangle \,.
\end{eqnarray}

Up to the leading order of $v$, the matrix-elements of the following relevant NRQCD effective operators will be useful in our calculation  \cite{Beneke:2008pi}
\begin{subequations}\label{eq:3}
\begin{eqnarray}
\mathcal{O}(^1S_0)&\equiv& \bar\psi_{bv}
\gamma_5 \chi_{cv}\,,\\[0.2cm]
\mathcal{O}^\mu(^3S_1)&\equiv&\bar\psi_{bv}
 \gamma_{\top}^{\mu} \chi_{cv}\,,\\
\mathcal{O}^\mu (^1P_1)&\equiv&
\bar\psi_{bv} \left[\left(-\frac{i}{2}\right)
\stackrel{\leftrightarrow}{D}_{\top}^{\mu}\gamma_5\right] \chi_{cv}\,, \\
\mathcal{O}(^3P_0)&\equiv&
\bar\psi_{bv} \left[-\frac{1}{\sqrt{3}}\left(-\frac{i}{2}\right)
\stackrel{\leftrightarrow}{\Slash D}_{\top}\right] \chi_{cv}\,, \\
\mathcal{O}^{\rho\mu\nu}(^3P_1)&\equiv&\frac{1}{2\sqrt{2}}
\bar\psi_{bv} \left(-\frac{i}{2}\right)
\stackrel{\leftrightarrow}{D}_{\top}^{\rho}\left[\gamma_\top^\mu,\gamma_{\top}^{\nu}
\right]\gamma_5 \chi_{cv}\,,\\
\mathcal{O}^\mu(^3P_1)&\equiv&\frac{1}{2\sqrt{2}}
\bar\psi_{bv} \left(-\frac{i}{2}\right)
\left[\stackrel{\leftrightarrow}{\Slash D}_{\top},\gamma_{\top}^{\mu}
\right]\gamma_5 \chi_{cv}\,,\\
\mathcal{O}^{\mu\nu}(^3P_2)&\equiv&
\bar\psi_{bv} \left[\left(-\frac{i}{2}\right)
\stackrel{\leftrightarrow}{D}_{\top}^{(\mu} \gamma_{\top}^{\nu)}\right] \chi_{cv} \,. 
\end{eqnarray}
\end{subequations}
Here $\stackrel{\leftrightarrow}{D}^\mu=\stackrel{\rightarrow}{D}^\mu-\stackrel{\leftarrow}{D}^\mu=\stackrel{\rightarrow}{\partial}^\mu
-\stackrel{\leftarrow}{\partial}^\mu-2ig_sA^\mu$, and $a_{\top}^{(\mu} b_{\top}^{\nu)}=(a_\top^\mu b_\top^\nu+a_\top^\nu b_\top^\mu)/2-a_\top\cdot b_\top(g^{\mu\nu}-v^\mu v^\nu)/(d-1) $ with $d=4$ means the symmetric 3-D traceless part of rank-2 tensor $a_\top^\mu b_\top^\nu$.

Using the spin symmetry of heavy quark system to relate the various matrix-elements of S-wave operators and P-wave operators, we have
\begin{eqnarray}\label{eq:OP}
\left\{\begin{array}{rcl}
\langle B_c(^1S_0) |\mathcal{O}(^{1}S_0) |0 \rangle
&=&  \langle \mathcal{O}(^{1}S_0) \rangle,
\\
\langle B_c(^3S_1)|\mathcal{O}^\mu(^{3}S_1) |0\rangle
&=& \varepsilon^{*\mu} \, \langle \mathcal{O}(^{1}S_0) \rangle,\\
\langle B_c(^1P_1) |\mathcal{O}^\mu(^{1}P_1) |0 \rangle
&=&  \varepsilon^{*\mu} \,\langle \mathcal{O}(^{3}P_0) \rangle,
\\
\langle B_c(^3P_0) |\mathcal{O}(^{3}P_0) |0\rangle
&=&  \langle \mathcal{O}(^{3}P_0) \rangle,\\
\langle B_c(^3P_1) |\mathcal{O}^{\rho\mu\nu} (^{3}P_1) |0\rangle
&=& -\frac{1}{2}\left(\varepsilon^{*\mu}\left(g^{\rho\nu}-\frac{P^\rho P^{\nu}}{m_{3A}^2}\right)-\varepsilon^{*\nu}\left(g^{\rho\mu}-\frac{P^\rho P^{\mu}}{m_{3A}^2}\right) \right)\, \langle \mathcal{O}(^{3}P_0) \rangle,
\\
\langle B_c(^3P_1) |\mathcal{O}^\mu (^{3}P_1) |0\rangle
&=& \varepsilon^{*\mu} \, \langle \mathcal{O}(^{3}P_0) \rangle,
\\
\langle B_c(^3P_2) |\mathcal{O}^{\mu\nu}(^{3}P_2) |0\rangle
&=&  \varepsilon^{*\mu\nu} \,
 \langle \mathcal{O}(^{3}P_0) \rangle\,.
\end{array}\right.
\end{eqnarray}
At the leading order of $\alpha_s$ and $v$, these NRQCD matrix-elements can be related to the Schr\"odinger wave functions at the origin 
within the color-singlet model  \cite{Beneke:2008pi}
\begin{subequations}\label{metowf}
\begin{eqnarray}
\langle\mathcal{O}(^{1}S_0)\rangle &=&\sqrt{2 N_c} \sqrt{2M_P}
\,\sqrt{\frac{1}{4\pi}} R_{10}(0)\,,
\\
\langle\mathcal{O}(^{3}P_0)\rangle &=&\sqrt{2 N_c} \sqrt{2M_{S}}
\,(-i)\,\sqrt{\frac{3}{4\pi}} R^\prime_{21}(0)\,.
\end{eqnarray}
\end{subequations}
Here $N_c=3$ is the color number, $M_P$ and $M_{S}$ are the masses for $^1S_0$ and $^3P_0$ states of $B_c$ mesons, respectively, $R_{nl}(r)$ denotes the radial Schr\"{o}dinger wave function of the $B_c$ meson with radial quantum number $n$ and orbit-angular momentum $l$, the prime denotes a derivative with respect of $r$.

\subsection{Tree-level matching}

In this subsection, we explain how to extract the short-distance coefficient at tree-level. This extraction is done through matching the matrix-elements between vacuum and a pair of $b$ quark and anti-$c$ quark state in which heavy quarks move non-relativistically in their center of mass frame. Once the matching at tree-level is done, the extension to NLO is natural.

First we set the momenta for $b$-quark and $\bar c$-quark with non-relativistic relative motion as
\begin{eqnarray}
p_b^\mu=m_b v^\mu+q^\mu\,,~~p_c^\mu=m_c v^\mu+\tilde{q}^\mu\,,~~p_b^2=m_b^2\,,~~p_c^2=m_c^2\,,
\end{eqnarray}
where  $q^\mu$ and $\tilde{q}^\mu$ are the residual momenta of $b$ quark and anti-$c$ quark, respectively, satisfying $\tilde{q}_\top^\mu=-q_\top^\mu\sim {\cal O}(v)$ and $v\cdot q\sim v\cdot \tilde{q}\sim {\cal O}(v^2)$.
The total momentum of heavy quark pair
\begin{eqnarray}
P^\mu=p_b^\mu+p_c^\mu=m_H v^\mu\,,~~m_H\equiv \sqrt{P^2}\approx M+{\cal O}(v^2)\,,	
\end{eqnarray}
where $M\equiv m_b+m_c$. 

  By use of equations of motion and the power-counting for the residual momenta, the on-shell spinors of quark and anti-quark can be expanded in $v$ as
\begin{subequations}
\begin{eqnarray}
 u_b(p_b)&=&\left(1+\frac{\slash{q}_\top}{2m_b}+{\cal O}(v^2)\right)u_{bv}(p)=\left(1+\frac{\slash{\bar q}_\top}{2m_b}+{\cal O}(v^2)\right)u_{bv}(p)\,,~~~\\
 v_c(p_c)&=&\left(1-\frac{\slash{\tilde{q}}_\top}{2m_c}+{\cal O}(v^2)\right)v_{cv}(p)=\left(1+\frac{\slash{\bar q}_\top}{2m_c}+{\cal O}(v^2)\right)v_{cv}(p)\,,
\end{eqnarray}
\end{subequations}
where $\bar q_\top^\mu\equiv \frac{(q-\tilde{q})_\top^\mu}{2}$ and 
\begin{eqnarray}
u_v(p)=\frac{1+\slash{v}}{2}u(p)\,,~~v_v(p)=\frac{1-\slash{v}}{2}v(p)\,.
\end{eqnarray}

Thus, we have the matrix-element at tree-level
\begin{eqnarray}\label{eq:17}
&&\langle  b^a(p_b)\bar c^b(p_c)| Q[\Gamma](x)| 0  \rangle = \delta^{ab}\int\frac{d \omega}{2\pi}e^{-i(x-n_+p_b/n_+P)\omega n_+ P}\bar u_b(p_b)\slash{n}_+\Gamma v_c(p_c)\nn\\
&=&\frac{\delta^{ab}}{n_+P}\delta\left(x-\frac{n_+p_b}{n_+P}\right)\bar u_b(p_b)\slash{n}_+\Gamma v_c(p_c) \nn\\
&=&\frac{\delta^{ab}}{n_+P}\Bigg[ \Big( \delta(x-x_0)-\delta^\prime(x-x_0)\frac{n_+\bar q}{n_+P} \Big)\bar u_{bv}(p_b)\slash{n}_+\Gamma v_{cv}(p_c)\nn\\
&&+\frac{\delta(x-x_0)}{4m_r}\left(\bar u_{bv}(p_b)\{\slash{\bar q},\slash{n}_+\Gamma \}v_{cv}(p_c)+(1-2 x_0)\bar u_{bv}(p_b)[\slash{\bar q},\slash{n}_+\Gamma]v_{cv}(p_c)\right)+\mathcal{O}(v^2) \Bigg]\,,\nn\\
\end{eqnarray}
where $x_0\equiv m_b/(m_b+m_c)$, $m_r\equiv\frac{m_bm_c}{m_b+m_c}=x_0\bar x_0 M$ is the reduced mass of  the quark-anti-quark pair, and $a$, $b$ are color indices for the quark and anti-quark, respectively. Note that (\ref{eq:17}) has an extra term proportional to the spin structure $[\slash{\bar q},\slash{n}_+\Gamma]$ compared with (2.43) in \cite{Wang:2013ywc}. This extra term will make some LCDAs for the P-wave $B_c$ mesons more complicated than the ones for quarkonia. 

By the matching equation (\ref{eq:matching}) and inserting $\Gamma=1\,,\gamma_5\,,\gamma^\alpha_\perp$ and $\gamma_\perp^\alpha\gamma_5$,  we can get the distribution parts of LCDAs and decay constants  at tree-level
\begin{eqnarray}
    && \left\{ \begin{array}{l} \hat \phi_P^{(0)}(x)=\phi_V^{\parallel(0)}(x)=\phi_V^{\perp(0)}(x)=\delta(x-x_0) \\\\
   f_P^{(0)}=f_V^{(0)}=f_V^{\perp(0)}=\frac{i}{M}\langle   \mathcal{O}(^1S_0)   \rangle   \end{array}  \right.\,, 
 \end{eqnarray}
for the S-wave $B_c$ mesons,
\begin{eqnarray}
&&\left\{\begin{array}{l} \hat \phi_{1A}^{\parallel(0)}(x)=\delta(x-x_0)+\frac{2x_0\bar x_0}{2x_0-1}\delta^\prime(x-x_0) \\\\
   f_{1A}^{(0)}=i\frac{1-2x_0}{2x_0\bar x_0 M^2}\langle \mathcal{O}(^3P_0) \rangle \end{array}\right. \,, ~~\left\{ \begin{array}{l} \hat{\phi}_{1A}^{\perp(0)}(x)=\delta(x-x_0)  \\\\
f_{1A}^{\perp(0)}=-\frac{i}{2x_0\bar x_0 M^2}\langle \mathcal{O}(^3P_0) \rangle \end{array}   \right.   \,,
\end{eqnarray}
for the $B_c$ meson in $^1P_1$ state,
\begin{eqnarray}&&
\left\{\begin{array}{l}  \hat \phi^{(0)}_S(x)=\delta(x-x_0)+\frac{2x_0\bar x_0}{3(2x_0-1)}\delta^\prime(x-x_0) \\\\
  f_S^{(0)}=-\frac{(2x_0-1)\sqrt{3}}{2x_0\bar x_0 M^2}\langle \mathcal{O}(^3P_0) \rangle   \end{array}\right.\,, \end{eqnarray}
for the $B_c$ meson in $^3P_0$ state,
\begin{eqnarray}
    &&\left\{ \begin{array}{l}\hat \phi_{3A}^{\parallel(0)}(x)=\delta(x-x_0) \\\\
   f_{3A}^{(0)}=i\frac{\sqrt{2}}{2x_0\bar x_0 M^2}\langle   \mathcal{O}(^3P_0)   \rangle  \end{array}  \right. \,,~~\left\{\begin{array}{l}  \hat{\phi}_{3A}^{\perp(0)}(x)=\delta(x-x_0)+\frac{x_0\bar x_0}{(2x_0-1)}\delta^\prime(x-x_0)  \\\\
f_{3A}^{\perp(0)}=i\frac{\sqrt{2}(2x_0-1)}{2x_0\bar x_0M^2}\langle \mathcal{O}(^3P_0) \rangle \end{array}\right.\,, \end{eqnarray}
for the $B_c$ meson in $^3P_1$ state,  and
\begin{eqnarray}&&
\left\{ \begin{array}{l} \hat \phi_T^{\parallel(0)}(x)= \hat \phi_T^{\perp(0)}(x)=-\delta^\prime(x-x_0)/2 \\\\
f^{(0)}_T=f^{\perp(0)}_T=\frac{2}{M^2}\langle  \mathcal{O}(^3P_0)\rangle \end{array}  \right.\,,
\end{eqnarray}
for the $B_c$ meson in $^3P_2$ state. Here the superscript $(0)$ denotes the quantities which are calculated at the leading order of $\alpha_s$. 

By setting $m_b=m_c=m$, i.e. $x_0=1/2$, we recover the results $f\hat \phi(x)$ for quarkonia at tree-level given in \cite{Wang:2013ywc}.

\section{The calculations of the LCDAs at NLO\label{sect:NLO}}

At the NLO of $\alpha_s$ and in the dimensional regularization (DR) scheme, the bare matrix-element of $Q[\Gamma](x)$ in the Feynman gauge is written as\footnote{Here we set the momentum of gluon in the loop as $k-\bar q$ as in \cite{Beneke:1997zp}.  And we have simplified the spin-structures in the last terms of (\ref{eq:bare}) by  considering the fact that $\Gamma$ is either commuting or anti-commuting with $\slash {n}_+$. }
\begin{eqnarray} \label{eq:bare}
&& \langle  b^a(p_b)\bar c^b(p_c) |Q[\Gamma](x)   |0    \rangle^{\rm bare}
\nonumber \\
&=&\delta^{ab}\delta\left(x-\frac{n_+p_b}{n_+P}\right)\frac{\bar u_b(p_b)\slash{n}_+\Gamma v_c(p_c)}{n_+P}\nn\\
   &&+\frac{\alpha_s}{4\pi}C_F\delta^{ab}\int [dk]\frac{\bar u_b(p_b)\gamma^\mu(\slash{k}-\slash{\bar q}+\slash{p}_b+m_b)\slash{n}_+\Gamma(\slash{k}-\slash{\bar q}-\slash{p}_c+m_c)\gamma_\mu v_c(p_c)}{n_+P[(k-\bar q)^2+i\epsilon][(k+p_b-\bar q)^2-m_b^2+i\epsilon][(k-p_c-\bar q)^2-m_c^2+i\epsilon]}\nonumber\\
&&~~~~~~~~~~~~~~~~~~~\times\delta\left(x-\frac{n_+(p_b+k-\bar q)}{n_+P}\right)\nn\\
    &&-\frac{\alpha_s}{4\pi}C_F\delta^{ab}\int [dk]\frac{2 n_+(k-\bar q+p_b)\bar u_b(p_b)\slash{n}_+\Gamma v_c(p_c)}{n_+P[n_+(k-\bar q)][(k-\bar q)^2+i\epsilon][(k+p_b-\bar q)^2-m_b^2+i\epsilon]}\nonumber\\
&&~~~~~~~~~~~~~~~~~~~\times\left[ \delta\left(x-\frac{n_+(k+p_b-\bar q)}{n_+P}\right)-\delta\left(x-\frac{n_+p_b}{n_+P}\right) \right]\nn\\
   &&-\frac{\alpha_s}{4\pi}C_F\delta^{ab}\int [dk]\frac{2n_+ (k-\bar q-p_c)\bar u_b(p_b)\slash{n}_+\Gamma v_c(p_c)}{n_+P[n_+(k-\bar q)][(k-\bar q)^2+i\epsilon][(k-p_c-\bar q)^2-m_c^2+i\epsilon]}\nonumber\\
&&~~~~~~~~~~~~~~~~~~~\times\left[ \delta\left(x-\frac{n_+(k+p_b-\bar q)}{n_+P}\right)-\delta\left(x-\frac{n_+p_b}{n_+P}\right) \right]\,.
\end{eqnarray}
Here $\alpha_s=g_s^2/(4\pi)$ is the running strong coupling, $C_F=(N_c^2-1)/2N_c$ with $N_c=3$ is rank-2 Casimir in the fundamental representation of SU(3) group, and 
\begin{eqnarray}
  [dk]\equiv \frac{(4\pi)^2}{i} \left(\frac{e^{\gamma_{\rm E}}\mu^2}{4\pi}\right)^\varepsilon \frac{dn_+k d^{d-2}k_\perp dn_-k}{2(2\pi)^d}\,,
\end{eqnarray}
with $\gamma_{\rm E}=0.5772...$ being the Euler constant, and $d=4-2 \varepsilon$ the dimension of the space-time where the extra dimension $\varepsilon$ is used to regulate both of the ultraviolet and infrared divergences appearing in our calculation.
The loop momentum $k$ has been decomposed into
\begin{eqnarray}
  k^\mu=n_+k\frac{n_-^\mu}{2}+n_-k\frac{n_+^\mu}{2}+k_\perp^\mu\,,
\end{eqnarray}
where we set $n_{\pm}$ within 4 dimensions, and $k_\perp$ runs over the remaining $d-2$ dimensions. 

\subsection{Matching procedure by method of threshold expansion}

The standard procedure to extract the short-distance coefficient $C_\Gamma^n(x,\mu)$ in (\ref{eq:matching}) at the NLO of $\alpha_s$ is to calculate the matrix-elements of $Q[\Gamma](x,\mu)$ and $O_{\Gamma,n}^{\rm NRQCD}$ at NLO  separately, and then match them together, as what done in \cite{Ma:2006hc}. However, in this paper, we will perform the matching procedure by adopting the  threshold expansion developed in \cite{Beneke:1997zp}, as in \cite{Wang:2013ywc,Bell:2008er}.

In the threshold expansion, the $v$-expansion of full QCD loop integration is reproduced by a sum of several loop integrations over different loop momentum regions which are characterized by the different power-countings under $v$-expansion. 
Since our definitions of the LCDAs are boost-invariant, we choose $v^\mu=(1,\vec{0})$, i.e. the quark and anti-quark pair is in its rest frame. Thus, the most important integration regions which contribute to the full loop integrals in (\ref{eq:bare}) are \cite{Beneke:1997zp}: 
\begin{eqnarray}
\left\{\begin{array}{rl}
\text{hard region:} &~~k^\mu \sim {\cal O}(M),\\
\text{soft region:}&~~k^\mu \sim {\cal O}(Mv),\\
\text{ultra-soft region:}&~~k^\mu  \sim {\cal O}(Mv^2),\\
\text{potential region:}&~~v\cdot k \sim {\cal O}(Mv^2),~k^\mu_\top \sim {\cal O}(Mv).
\end{array}\right.
\end{eqnarray}
As commonly known, the contributions from soft, ultra-soft and potential regions reproduce the loop corrections to the matrix-element of $O_{\Gamma,n}^{\rm NRQCD}$  in NRQCD, and the contribution from the hard region gives to the loop corrections to the short-distance coefficient $C_\Gamma^n(x,\mu)$.  Thus, the threshold expansion can simplify the matching procedure greatly. We only need to calculate the hard integration to extract the short-distance coefficients $C_\Gamma^n(x,\mu)$.

In order to get the results up to ${\cal O}(v)$, we need the following $v$-expansions of the loop integrands in the hard region. For examples, 
\begin{eqnarray}
&&\frac{1}{(k-\bar q)^2+i\epsilon}=\frac{1}{k^2+i\epsilon}\left(1-\frac{2\bar q\cdot k}{k^2+i\epsilon}+{\cal O}(v^2)\right)\,,\nn\\
&&\frac{1}{n_+(k-\bar q)}=\frac{1}{n_+k}\left(1+\frac{n_+\bar q}{n_+k}+{\cal O}(v^2)\right)\,,\nn\\
&&\frac{1}{(k+p_b-\bar q)^2-m_b^2+i\epsilon}=\frac{1}{k^2+2 m_bv\cdot k+i\epsilon}+{\cal O}(v^2)\,,\nn\\
&&\frac{1}{(k-p_c-\bar q)^2-m_c^2+i\epsilon}=\frac{1}{k^2-2 m_cv\cdot k+i\epsilon}+{\cal O}(v^2)\,,\nn
\end{eqnarray}
for the denominators of the integrands, and
\begin{eqnarray}
   &&\bar u_{b}(p_b)\gamma^\mu (\slash{k}-\slash{\bar q})\slash{n}_+\Gamma (\slash{k}-\slash{\bar q})\gamma_\mu v_c(p_c) \nn\\
    &=&\bar u_{bv}(p_b)\Big(1 +\frac{\slash{\bar q}}{2m_b} \Big)\gamma^\mu(\slash{k}-\slash{\bar q})\slash{n}_+\Gamma(\slash{k}-\slash{\bar q})\gamma_\mu\Big(1+ \frac{\slash{\bar q}}{2m_c}\Big)v_{cv}(p_c)+{\cal O}(v^2)\nn\\
   &=& \bar u_{bv}(p_b)\Big(\gamma^\mu\slash{k}\slash{n}_+\Gamma \slash{k}\gamma_\mu-\gamma^\mu\slash{\bar q}\slash{n}_+\Gamma \slash{k}\gamma_\mu-\gamma^\mu\slash{k}\slash{n}_+\Gamma \slash{\bar q}\gamma_\mu\nn\\
&&~~~~~~~~~+\frac{1}{2m_b}\slash{\bar q}\gamma^\mu\slash{k}\slash{n}_+\Gamma \slash{k}\gamma_\mu+\frac{1}{2m_c}\gamma^\mu\slash{k}\slash{n}_+\Gamma \slash{k}\gamma_\mu\slash{\bar q}\Big)v_{cv}(p_c)+{\cal O}(v^2)\, ,\nn
    \end{eqnarray}
for the numerators of the integrands. The delta-functions appearing in  (\ref{eq:bare}) should be also expanded in $v$. 

After the expansion, a general hard loop integration will look like
\begin{eqnarray}
\int[dk]\frac{f(n_+k)\times (1,k_\perp^2, n_-k)}{[k^2+i\epsilon]^a[k^2+2m_bv\cdot k+i\epsilon]^b[k^2-2m_c v\cdot k+i\epsilon]^c}\,,\nn
\end{eqnarray}
which can be done straightforwardly by integrating over $n_-k$ first by use of residue theorem,  and then integrating over $k_\perp$. The final result of the integral is an integration over $n_+k$ within $[-m_b n_+v,m_c n_+v]$, which will be translated to an integration over a light-cone fraction, say $y$, within $[0,1]$ as in \cite{Wang:2013ywc}.

\subsection{Simplifications of the spin structures and schemes on $\gamma_5$}

After the tedious expansions of integrands in hard region, we get various complicated spinor bilinears with complicated spin-structures like 
\begin{eqnarray}
\bar u(p_1)\cdots \slash n_+\Gamma\cdots v(p_2)\,.
\end{eqnarray} 
To simplify them further, we have to fix the scheme to treat $\gamma_5$ in DR. There are two widely-used schemes about $\gamma_5$ in DR, one is the naive dimensional regularization (NDR) scheme  \cite{Chanowitz:1979zu}, in which 
 $\{\gamma_5,\gamma^\mu\}=0$, $\{\gamma^\mu,\gamma^\nu\}=2 g^{\mu\nu}$ and $g_\mu^\mu=d$; 
the other is the t'Hooft-Veltman (HV) scheme \cite{'tHooft:1972fi,Breitenlohner:1977hr}, in which $\gamma_5\equiv i\gamma^0\gamma^1\gamma^2\gamma^3$, and $\{\gamma^\mu,\gamma_5\}=0$ for $\mu=0,1,2,3$ but $[\gamma^\mu,\gamma_5]=0$ for $\mu=4,...,d-1$. 
In this paper, we will compute the NLO corrections to the LCDAs in both the NDR and HV schemes. 

To simplify the spinor bilinears, we try to use only identities which hold in both NDR and HV schemes, such as $\{\gamma^\mu,\gamma^\nu\}=2g^{\mu\nu}$, $\{\slash n_{\pm},\Gamma\}=0$ or $[\slash n_{\pm},\Gamma]=0$ and on-shell conditions for the external spinors as much as possible. In the end, it turns out that the only possible $\gamma_5$-dependent  structures are 
\begin{eqnarray}\label{eq:gamma}
\gamma^\rho \slash n_+\Gamma\gamma_\rho~{\rm and }~\gamma^\rho \gamma^\sigma\slash n_+\Gamma\gamma_\sigma\gamma_\rho\,.
\end{eqnarray}
With $\Gamma=1\,,\gamma_5\,,\gamma_\perp^\alpha$ and $\gamma_\perp^\alpha \gamma_5$ ({note that the index $\alpha_\perp$ runs within 4-dimensions since it is from the definition of the LCDAs), we find that $\gamma^\rho \slash n_+\Gamma \gamma_\rho \equiv c_{\slash n_+\Gamma} \slash n_+\Gamma$ where
\begin{eqnarray}
c_{\slash n_+\Gamma}=\left\{
\begin{array}{ll}
2-d\,, &~~~\Gamma=\mathbf{1}\,,\\
d-2\,, &~~~\Gamma=\gamma_5\,,\\
d-4\,, &~~~\Gamma=\gamma^\alpha_\perp\,,\\
4-d\,, &~~~\Gamma=\gamma^\alpha_\perp\gamma_5\,,
\end{array}
\right.
\end{eqnarray}
in the NDR scheme and 
\begin{eqnarray}
c_{\slash n_+\Gamma}=\left\{
\begin{array}{ll}
2-d\,, &~~~\Gamma=\mathbf{1}\,,\\
6-d\,, &~~~\Gamma=\gamma_5\,,\\
d-4\,, &~~~\Gamma=\gamma^\alpha_\perp\,,\\
d-4\,, &~~~\Gamma=\gamma^\alpha_\perp\gamma_5\,,
\end{array}
\right.
\end{eqnarray}
in the HV scheme. 

\subsection{Operator renormalization}

The hard part of the renormalized matrix-element reads
\begin{eqnarray}
&&\langle b^a(p_b)\bar c^b(p_c)\vert Q[\Gamma](x)\vert 0\rangle^{\rm ren}_{\rm hard}=\left(Z^{\rm os}_{b,2} Z^{\rm os}_{c,2}\right)^{1/2} \int_0^1 dy Z_{\slash n_+ \Gamma}(x,y)\langle Q^a(p_b)\bar Q^b(p_c)\vert Q[\Gamma](y)\vert 0\rangle^{\rm bare}_{\rm{hard}}\,,\nn\\
\end{eqnarray}
where the on-shell renormalization constant for the heavy quark is
\begin{eqnarray}
  Z_{Q,2}^{\rm os}=1-\frac{\alpha_s}{4\pi} C_F\left( \frac{3}{\varepsilon}+3\ln\frac{\mu^2}{m_Q^2}+4 \right)\,,
\end{eqnarray}
with $Q=b$ or $c$, and the renormalization kernels for the operator $Q[\Gamma](x)$ in the $\overline{\textrm{MS}}$ scheme are
\begin{subequations}
\begin{eqnarray}
Z_{\slash n_+\gamma_5}(x,y)&=&Z_{\slash n_+}(x,y)\,=\,\delta(x-y)-\frac{\alpha_s}{4\pi}C_F \frac{2}{\varepsilon} V_0(x,y)\,,\\
Z_{\slash n_+\gamma_\perp^\alpha\gamma_5}(x,y)&=&Z_{\slash n_+ \gamma_\perp^\alpha}(x,y)\,=\,\delta(x-y)-\frac{\alpha_s}{4\pi}C_F \frac{2}{\varepsilon} V_\perp(x,y)\,,
\end{eqnarray}
\end{subequations}
with the Brodsky-Lepage kernel being
\begin{subequations}
\begin{eqnarray}
V_0(x,y)&=&\left[\frac{\bar x}{\bar 
y}\left(1+\frac{1}{x-y}\right)\theta(x-y)+\frac{x}{y}\left(1+\frac{1}{y-x}\right)
\theta(y-x)\right]_+\,,\\
V_\perp(x,y)&=&V_0(x,y)-\left[\frac{1- x}{1-
y}\theta(x-y)+\frac{x}{y} \theta(y-x)\right]\nn\\
&=&\left[\frac{\bar x}{\bar 
y}\frac{1}{x-y}\theta(x-y)+\frac{x}{y}\frac{1}{y-x}
\theta(y-x)\right]_+-\frac{1}{2}\delta(x-y)\,.
\end{eqnarray}
\end{subequations}
In total, the hard part of the renormalized matrix-element can be expanded up to ${\cal O}(v)$ as
\begin{eqnarray}
&&\langle b^a(p_b) \bar c^b(p_c)\vert Q[\Gamma](x)\vert 0\rangle^{\rm ren}_{\rm hard} \nn \\
&=&\delta^{ab}\frac{\bar u_{bv}(p_b)\slash n_+\Gamma v_{cv}(p_c)}{n_+P}S_0+\frac{\delta^{ab}}{4m_r}\frac{\bar u_{bv}(p_b)\left\{\slash{\bar  q},\slash n_+\Gamma\right\} v_{cv}(p_c)}{n_+P}S_1\nn\\
&&+\frac{\delta^{ab}}{4m_r}(1-2x_0)\frac{\bar u_{bv}(p_b)\left[\slash {\bar  q},\slash n_+\Gamma\right] v_{cv}(p_c)}{n_+P} S_2\nn\\
&&+\delta^{ab}\frac{n_+\bar q}{n_+P}\frac{\bar u_{bv}(p_b)\slash n_+\Gamma v_{cv}(p_c)}{n_+P}S_3+{\cal O}(v^2)\,,
\end{eqnarray}
where $S_i$ ($i=0,1,2,3$) are both ultraviolet and infrared finite distributions over the light-cone fraction $x$.  Eventually, after substituting $\Gamma$ with $1,\gamma_5,\gamma_\perp^\alpha$ and $\gamma_\perp^\alpha\gamma_5$,  one can extract all the three LCDAs for the S-wave states $f_P \hat\phi_P(x;\mu)$, $f_V\hat \phi_V^\parallel(x;\mu)$ and $f_V^\perp \hat \phi_V^\perp(x;\mu)$ from $S_0$, two of the LCDAs for the P-wave states $f_{1A}^\perp \hat \phi_{1A}^\perp(x;\mu)$ and $f_{3A} \hat \phi_{3A}^\parallel(x;\mu)$ from $S_1$,  two LCDAs for the $^3P_2$ states $f_T\hat \phi_T^\parallel(x;\mu)$ and $f_T^\perp \phi_T^\perp(x;\mu)$ from $S_3$, and the rest three LCDAs for the P-wave states, $f_{1A}\hat \phi_{1A}^\parallel(x;\mu)$, $f_S\hat\phi _S(x;\mu)$ and $f_{3A}\hat \phi_{3A}^\perp (x;\mu)$ from both $S_2$ and $S_3$.

\subsection{Final results for the LCDAs at  NLO}
Here we present the final results for the LCDAs at the NLO of $\alpha_s$ and leading order of $v$. The three LCDAs for the S-wave $B_c$ mesons are
\begin{subequations}
\begin{eqnarray}
 \hat \phi_P(x;\mu)&=&\delta(x-x_0)
+\frac{\alpha_s}{4\pi}C_F \Bigg\{\Phi_1(x,x_0)
+8\Delta\left[\frac{x}{x_0}\theta(x_0-x)+\left( \begin{array}{l} x\leftrightarrow \bar x\\ x_0\leftrightarrow \bar x_0  \end{array} \right)\right]_{+}\Bigg\} \,, \nn\\\\
   \hat \phi_V^\parallel(x;\mu)&=& \delta(x-x_0)
+\frac{\alpha_s}{4\pi}C_F \Bigg\{\Phi_1(x,x_0)
-4\left[\frac{x}{x_0}\theta(x_0-x)+\left( \begin{array}{l} x\leftrightarrow \bar x\\ x_0\leftrightarrow \bar x_0  \end{array} \right)\right]_{+}\Bigg\} \,, \label{eq:INNRQCD}\\
 \hat \phi_V^\perp(x;\mu)&=&\delta(x-x_0)
+\frac{\alpha_s}{4\pi}C_F \Bigg\{\Phi_1(x,x_0) \nn\\
&&~~
-2\left[\left( \ln{\frac{\mu^2}{M^2(x_0-x)^2}}-1\right)\left( \frac{x}{x_0}\theta(x_0-x)+\left( \begin{array}{l} x\leftrightarrow \bar x\\ x_0\leftrightarrow \bar x_0  \end{array} \right) \right) \right]_+\Bigg\} \,, 
\end{eqnarray}
\end{subequations}
with 
\begin{eqnarray}
\Phi_1(x,x_0)&=&2\left[ \left( \ln{\frac{\mu^2}{M^2(x_0-x)^2}}-1 \right)\left( \frac{x_0+\bar x}{x_0-x}\frac{x}{x_0}\theta(x_0-x)+\left( \begin{array}{l} x\leftrightarrow \bar x\\ x_0\leftrightarrow \bar x_0  \end{array} \right) \right) \right]_+ \nn\\
&&~~+\left[ \frac{4x\bar x}{(x_0-x)^2}\right]_{++}+\left[ 4x_0\bar x_0 \ln{\frac{x_0}{\bar x_0}}+2(2x_0-1) \right]\delta^\prime(x-x_0)\,,
\end{eqnarray}
and the corresponding decay constants are
\begin{subequations}
\begin{eqnarray}
  f_P&=&\frac{i}{M}\langle   \mathcal{O}(^1S_0)   \rangle \left\{ 1+\frac{\alpha_s}{4\pi}C_F\left[-6+4\Delta+3(x_0-\bar x_0)\ln{\frac{x_0}{\bar x_0}}  \right] \right\}\,,\\
  f_V&=&\frac{i}{M}\langle  \mathcal{O}(^1S_0)\rangle \left\{ 1+ \frac{\alpha_s}{4\pi}C_F\left[ -8+3(x_0-\bar x_0)\ln{\frac{x_0}{\bar x_0}} \right] \right\}\,,\\
f_V^\perp&=&\frac{i}{M}\langle  \mathcal{O}(^1S_0)\rangle\left\{1+ \frac{\alpha_s}{4\pi}C_F \left[-\ln{\frac{\mu^2}{M^2}}-(3-8x_0)\ln x_0-(3-8\bar x_0)\ln \bar x_0-8 \right] \right\}\,. \nn\\
\end{eqnarray}
\end{subequations}
Here, $\Delta=0$ for the NDR scheme, and $\Delta=1$ for the HV scheme.

Similarly, the seven LCDAs for the P-wave $B_c$ mesons are
\begin{subequations}
 \begin{eqnarray}
   \hat \phi_{1A}^\parallel(x;\mu)&=& \hat \phi_{3A}^\parallel(x;\mu)+\frac{4x_0\bar x_0}{1-2 x_0}
 \hat \phi_T^\parallel(x;\mu)\nn\\
&&+\frac{4x_0\bar x_0}{1-2 x_0}\frac{\alpha_s}{4\pi}C_F \Bigg\{ -\left[\frac{x^2\theta(x_0-x)}{x_0^2(x_0-x)}-\frac{\bar x^2\theta(x-x_0)}{\bar x_0^2(x-x_0)}\right]_{++} \nn\\
&&~~~~~~-(2+4\Delta)\left[\frac{x\theta(x_0-x)}{x_0^2}-\frac{\bar x\theta(x-x_0)}{\bar x_0^2}\right]_{++}\nonumber\\
&&~~~~~~+\left[\frac{4}{3}\left(\ln\frac{\mu^2}{M^2}-2x_0\ln x_0-2\bar x_0\ln \bar x_0\right)+\frac{38-6\Delta}{9}\right]\delta^\prime(x-x_0)\nonumber \\
&&~~~~~~+\frac{1}{2}\left[  3(x_0-\bar x_0)\ln \frac{x_0}{\bar x_0}-4+4\Delta  -\frac{4x_0\bar x_0}{1-2x_0}\ln\frac{x_0}{\bar x_0}\right]\delta^\prime(x-x_0) \Bigg\} \,, \\
   \hat{\phi}_{1A}^{\perp}(x;\mu)&=&\delta(x-x_0)+\frac{\alpha_s}{4\pi}C_F \Bigg\{\Phi_2(x,x_0) \nn\\
&&~~~~~~~~~
-2\left[\left( \ln{\frac{\mu^2}{M^2(x_0-x)^2}}-1\right)\left( \frac{x}{x_0}\theta(x_0-x)+\left( \begin{array}{l} x\leftrightarrow \bar x\\ x_0\leftrightarrow \bar x_0  \end{array} \right) \right) \right]_+\Bigg\} \,, \\
 \hat \phi_S(x;\mu)&=&\delta(x-x_0)+\frac{\alpha_s}{4\pi}C_F \Phi_2(x,x_0)+\frac{4x_0\bar x_0}{1-2x_0}\frac{1}{3}\hat\phi_T^\parallel(x\,;\mu)
\nn\\
&&+\frac{4x_0\bar x_0}{1-2x_0}\frac{1}{3}\frac{\alpha_s}{4\pi}C_F \Bigg\{
-3\left[\frac{x(4x_0-x)\theta(x_0-x)}{x_0^2(x_0-x)}-\left( \begin{array}{l} x\leftrightarrow \bar x\\ x_0\leftrightarrow \bar x_0  \end{array} \right)\right]_{++}\nonumber\\
&&~~~~~~+\left[\frac{4}{3}\left(\ln\frac{\mu^2}{M^2}-2x_0\ln x_0-2\bar x_0\ln \bar x_0\right)-\frac{1}{9}\right]\delta^\prime(x-x_0)\nonumber \\
&&~~~~~~+\frac{1}{2}\left[  3(x_0-\bar x_0)\ln \frac{x_0}{\bar x_0}-2  -\frac{12x_0\bar x_0}{1-2x_0}\ln\frac{x_0}{\bar x_0}\right]\delta^\prime(x-x_0)\Bigg\}  \,,\\
\hat \phi_{3A}^\parallel(x;\mu)&=&\delta(x-x_0)
+\frac{\alpha_s}{4\pi}C_F \Bigg\{\Phi_2(x,x_0)
+(8\Delta-4)\left[\frac{x}{x_0}\theta(x_0-x)+\left( \begin{array}{l} x\leftrightarrow \bar x\\ x_0\leftrightarrow \bar x_0  \end{array} \right)\right]_{+}\Bigg\}  \,, \nn\\\\
  \hat{\phi}_{3A}^{\perp}(x;\mu)&=&\hat{\phi}_{1A}^{\perp}(x;\mu)+\frac{2x_0\bar x_0}{(1-2x_0)}\hat{\phi}_{T}^{\perp}(x;\mu)\nn\\&&
+\frac{2x_0\bar x_0}{(1-2x_0)}\frac{\alpha_s}{4\pi}C_F \Bigg\{-4\left[\frac{x\theta(x_0-x)}{x_0(x_0-x)}-\frac{\bar x\theta(x-x_0)}{\bar x_0(x-x_0)}\right]_{++}\nn\\
   &&	 +\left[\ln\frac{\mu^2}{M^2}-\frac{3-2 x_0}{2}\ln x_0-\frac{1+2 x_0}{2}\ln \bar x_0+1-\frac{4x_0\bar x_0}{1-2x_0}\ln\frac{x_0}{\bar x_0}\right]\delta^\prime(x-x_0)\Bigg\}\,, \nn\\\\
  \hat \phi_T^\parallel(x;\mu)&=&-\delta^\prime(x-x_0)/2+\frac{\alpha_s}{4\pi}C_F \Bigg\{-\left[ \ln \frac{\mu^2}{M^2(x_0-x)^2}\left(\frac{x}{x_0^2}\theta(x_0-x) -\left( \begin{array}{l} x\leftrightarrow \bar x\\ x_0\leftrightarrow \bar x_0  \end{array} \right)\right) \right]_{++}   \nn\\
&&~~~~~~ -\left[ \ln \frac{\mu^2}{M^2(x_0-x)^2}\left(\frac{x(2x_0-x)}{x_0^2(x_0-x)^2}\theta(x_0-x) -\left( \begin{array}{l} x\leftrightarrow \bar x\\ x_0\leftrightarrow \bar x_0  \end{array} \right)\right) \right]_{++}  \nn\\
   &&~~~~~~-2\left[\frac{\bar x_0 x}{(x_0-x)^3}\theta(x_0-x)-\left( \begin{array}{c} x\leftrightarrow \bar x\\ x_0\leftrightarrow \bar x_0  \end{array} \right) \right]_{+++}-x_0\bar x_0\ln \frac{x_0}{ \bar x_0}\delta^{\prime\prime}(x-x_0)\nn\\
   &&~~~~~~+\left[\frac{x(8x^2+2x_0(1+ x_0)-x(5+12x_0))}{2x_0^2(x_0-x)^2}\theta(x_0-x)-\left( \begin{array}{c} x\leftrightarrow \bar x\\ x_0\leftrightarrow \bar x_0  \end{array} \right)\right]_{++}\Bigg\}  \,,\nn\\\\
   \hat \phi_T^\perp(x;\mu)&=&-\delta^\prime(x-x_0)/2\nn\\
&&+\frac{\alpha_s}{4\pi}C_F \Bigg\{-\left[ \ln \frac{\mu^2}{M^2(x_0-x)^2}\left(\frac{x(2x_0-x)}{x_0^2(x_0-x)^2}\theta(x_0-x) -\left( \begin{array}{l} x\leftrightarrow \bar x\\ x_0\leftrightarrow \bar x_0  \end{array} \right)\right) \right]_{++}   \nn\\
   &&~~~~~~-2\left[\frac{\bar x_0 x}{(x_0-x)^3}\theta(x_0-x)-\left( \begin{array}{l} x\leftrightarrow \bar x\\ x_0\leftrightarrow \bar x_0  \end{array} \right) \right]_{+++}  -x_0\bar x_0  \ln \frac{x_0}{ \bar x_0}\delta^{\prime\prime}(x-x_0)\nn\\
   &&~~~~~~-2\left[\frac{x^2}{x_0^2(x_0-x)^2}\theta(x_0-x)-\left( \begin{array}{l} x\leftrightarrow \bar x\\ x_0\leftrightarrow \bar x_0  \end{array} \right)\right]_{++}\nn\\
   &&~~~~~~-\frac{(2x_0-1)}{2}\left[\frac{x(2x_0-x)}{x_0^2(x_0-x)^2}\theta(x_0-x)+\left( \begin{array}{l} x\leftrightarrow \bar x\\ x_0\leftrightarrow \bar x_0  \end{array} \right)\right]_{++}   \Bigg\}  \,,
\end{eqnarray}
\end{subequations}
with
\begin{eqnarray}
\Phi_2(x,x_0)&=&\Phi_1(x,x_0)-2x_0\bar x_0 \ln{\frac{x_0}{\bar x_0}}\delta^\prime(x-x_0) 
\nn\\&&~~~~~~
-2 \left[\frac{\bar x_0}{x_0}\frac{x(2 x_0-x)\theta(x_0-x)}{(x_0-x)^2}+\left( \begin{array}{l} x\leftrightarrow \bar x\\ x_0\leftrightarrow \bar x_0  \end{array} \right) \right]_{++}\,,
\end{eqnarray}
and the decay constants
\begin{subequations}
  \begin{eqnarray}
  f_{1A}&=&-i\frac{2x_0-1}{2x_0\bar x_0 M^2}\langle \mathcal{O}(^3P_0) \rangle\left\{ 1+\frac{\alpha_s}{4\pi}C_F\left[  3(x_0-\bar x_0)\ln \frac{x_0}{\bar x_0}-4+4\Delta  -\frac{4x_0\bar x_0}{1-2x_0}\ln\frac{x_0}{\bar x_0}\right]  \right\}\,,\nn\\\\
  f_{1A}^\perp&=&-\frac{i}{2x_0\bar x_0 M^2}\langle \mathcal{O}(^3P_0) \rangle \left\{1+ \frac{\alpha_s}{4\pi}C_F \left[-\ln{\frac{\mu^2}{M^2}}-(3-8x_0)\ln x_0-(3-8\bar x_0)\ln \bar x_0-4 \right] \right\}\,,\nn\\\\
 f_S&=&-\frac{(2x_0-1)\sqrt{3}}{2x_0\bar x_0 M^2}\langle \mathcal{O}(^3P_0) \rangle \left\{1+ \frac{\alpha_s}{4\pi}C_F\left[  3(x_0-\bar x_0)\ln \frac{x_0}{\bar x_0}-2  -\frac{12x_0\bar x_0}{1-2x_0}\ln\frac{x_0}{\bar x_0}\right] \right\}\,,\\
  f_{3A}&=&\frac{i\sqrt{2}}{2x_0\bar x_0 M^2} \langle \mathcal{O}(^3P_0) \rangle\left\{1+ \frac{\alpha_s}{4\pi}C_F\left[  3(x_0-\bar x_0)\ln \frac{x_0}{\bar x_0}-4+4\Delta  \right] \right\}\,, \\
      f_{3A}^\perp&=&i\frac{\sqrt{2}(2x_0-1)}{2x_0\bar x_0 M^2}\langle \mathcal{O}(^3P_0) \rangle\left\{ 1+\frac{\alpha_s}{4\pi}C_F \left[-\ln{\frac{\mu^2}{M^2}}-(3-8x_0)\ln x_0-(3-8\bar x_0)\ln \bar x_0-4\right.\right.\nonumber\\
&&~~~~\left.\left.-\frac{8x_0\bar x_0}{1-2x_0}\ln\frac{x_0}{\bar x_0}\right] \right\} \,,\\
  f_T&=&\frac{2}{M^2}\langle  \mathcal{O}(^3P_0)\rangle\left\{ 1-\frac{\alpha_s}{4\pi}C_F\left[\frac{8}{3}\left(\ln\frac{\mu^2}{M^2}-2 x_0 \ln  x_0-2 \bar x_0 \ln \bar x_0\right)+ \frac{88}{9}\right] \right\}\,,\\
  f_T^\perp&=&\frac{2}{M^2}\langle  \mathcal{O}(^3P_0)\rangle\left\{ 1-\frac{\alpha_s}{4\pi}C_F\left[3 \left(\ln\frac{\mu^2}{M^2}-2 x_0 \ln  x_0-2 \bar x_0 \ln \bar x_0\right)+10\right] \right\}
\,.
\end{eqnarray}
\end{subequations}
The $_{+++}$, $_{++}$, and $_{+}$ functions used in the above expressions are defined as
\begin{subequations}
\begin{eqnarray}
\int _0^1dx\Big[ f(x)  \Big]_{+++}g(x)&=&\int _0^1dx f(x)\left(g(x)-g(x_0)-g'(x_0)(x-x_0)-\frac{g''(x_0)}{2}(x-x_0)^2\right), \nn\\\\
\int _0^1dx\Big[ f(x)  \Big]_{++}g(x)&=&\int _0^1dx f(x)(g(x)-g(x_0)-g'(x_0)(x-x_0)),  \\
\int _0^1dx\Big[ f(x)  \Big]_{+}g(x)&=&\int _0^1dx f(x)(g(x)-g(x_0))\,.
\end{eqnarray}
\end{subequations}

And our results for $\hat \phi_M(x;\mu)$ preserve the normalizations in (\ref{condition}), and $f_M \hat \phi_M(x;\mu)$ obey the ERBL equations
\begin{eqnarray}
  \mu\frac{d}{d\mu}\left( f_M\hat \phi_M(x;\mu) \right)=\frac{\alpha_s(\mu)}{\pi} C_F \int_0^1 dy V_M(x,y) \left( f_M \hat\phi_M(y;\mu) \right)\,.
  \end{eqnarray}

We compare our results for the LCDAs of $B_c$ meson with those in \cite{Wang:2013ywc,Bell:2008er}. In \cite{Bell:2008er}, only $f_P\hat \phi_P(x;\mu)$ is calculated, and we find our result in the NDR scheme agrees with theirs. In \cite{Wang:2013ywc}, the authors calculated all ten LCDAs of quarkonia in both NDR and HV schemes. We find that our results agree with theirs when setting $m_b=m_c=m$.

\subsection{The inverse moments of the LCDAs}
In the practical applications of the leading twist LCDAs, since the lowest order hard-kernels $T_H(x)$ for many hard exclusive processes are in form of $1/x$ or $1/\bar x$, the inverse moments of the LCDAs are crucial for final amplitudes.
 We define the inverse moment of the LCDA as
\begin{eqnarray}
\left\langle \frac{1}{x}\right\rangle_\Gamma\equiv R_\Gamma\int_0^1 dx\frac{\hat\phi_\Gamma(x)}{x}\,,
\end{eqnarray}
with
 \begin{eqnarray}
   R_\Gamma \equiv \frac{f_\Gamma}{f_\Gamma^{(0)}}, ~~~\Gamma=P,V_\parallel,V_\perp,1A_\parallel,1A_\perp,S,3A_\parallel,3A_\perp,T_\parallel,T_\perp.
 \end{eqnarray}

 After implementing our results for the LCDAs at NLO, we get
\begin{subequations}
 \begin{eqnarray}
\left\langle \frac{1}{x}\right\rangle_P &=&\frac{1}{x_0}\left\{1+\frac{\alpha_s}{4\pi}C_F\left[\left(3+2\ln x_0\right)\ln\frac{\mu^2}{M^2}-3\ln\bar x_0-5\ln x_0-8\Delta\frac{x_0}{\bar x_0}\ln x_0 \right.\right.\nonumber\\
&&~~~~~~\left.\left.+4\,{\rm Li}_2(x_0)-2 \ln^2 x_0-\frac{2\pi^2}{3}\right]\right\} \,,\\
\left\langle \frac{1}{x}\right\rangle_V&=&\frac{1}{x_0}\left\{1+\frac{\alpha_s}{4\pi}C_F\left[\left(3+2\ln x_0\right)\ln\frac{\mu^2}{M^2}-3\ln\bar x_0-5\ln x_0+4\frac{x_0}{\bar x_0}\ln x_0\right.\right.\nonumber\\
&&\left.\left.~~~~~~  +4\,{\rm Li}_2(x_0)-2 \ln^2 x_0-\frac{2\pi^2}{3}\right]\right\} \,,\\
\left\langle \frac{1}{x}\right\rangle_{V_\perp}&=&\frac{1}{x_0}\left\{1+\frac{\alpha_s}{4\pi}C_F\left[\left(3+\frac{2}{\bar x_0}\ln x_0\right)\ln\frac{\mu^2}{M^2} -(3+4x_0)\ln\bar x_0-(5-4x_0)\ln x_0\right.\right.\nonumber\\
&&~~~~~~ \left.\left.-2\frac{x_0}{\bar x_0} \ln x_0+\frac{1}{\bar x_0}\left(4\,{\rm Li}_2(x_0)-2\ln^2 x_0-\frac{2\pi^2}{3}\right)\right]\right\} \,,\\ 
\left\langle \frac{1}{x}\right\rangle_{1A}&=&\frac{1}{x_0}\left\{1+\frac{\alpha_s}{4\pi}C_F\left[\left(3+2\ln x_0\right)\ln\frac{\mu^2}{M^2}-(1+2x_0)\ln\bar x_0-(5-2x_0)\ln x_0\right.\right.\nonumber\\
&&\left.\left.+(2-8\Delta)\frac{ x_0}{\bar x_0}\ln x_0 +4\,{\rm Li}_2(x_0)-2 \ln^2 x_0-\frac{2\pi^2}{3}+2-\frac{4x_0\bar x_0}{1-2x_0}\ln\frac{x_0}{\bar x_0}\right]\right\}\nonumber\\
&&-\frac{4x_0\bar x_0}{1-2x_0}\frac{1}{2x_0^2}\left\{1+\frac{\alpha_s}{4\pi}C_F\left[(1+2 \ln x_0)\ln\frac{\mu^2}{M^2}+\frac{3x_0-1}{\bar x_0}\ln x_0\right.\right.\nonumber\\
&&\left.\left.-(1-8\Delta)\frac{x_0^2}{\bar x_0^2}\ln x_0+4\,{\rm Li}_2(x_0)-2\ln^2 x_0-\frac{2 \pi ^2}{3}-(1-8\Delta)\frac{x_0}{\bar x_0}+3\right]\right\}\,,\\
\left\langle \frac{1}{x}\right\rangle_{1A_\perp}&=&\frac{1}{x_0}\left\{1+\frac{\alpha_s}{4\pi}C_F\left[\left(3+\frac{2}{\bar x_0}\ln x_0\right)\ln\frac{\mu^2}{M^2}-(1+6x_0)\ln\bar x_0-(5-6x_0)\ln x_0\right.\right.\nonumber\\
&&\left.\left.-4\frac{ x_0}{\bar x_0} \ln x_0+\frac{1}{\bar x_0}\left(4\,{\rm Li}_2(x_0)-2\ln^2 x_0-\frac{2\pi^2}{3}\right)+2\right]\right\} \,,\\ 
\left\langle \frac{1}{x}\right\rangle_{S}&=&\frac{1}{x_0}\left\{1+\frac{\alpha_s}{4\pi}C_F\left[\left(3+2\ln x_0\right)\ln\frac{\mu^2}{M^2}-(1+2x_0)\ln\bar x_0-(5-2x_0)\ln x_0\right.\right.\nonumber\\
&&\left.\left.-2\frac{ x_0}{\bar x_0}\ln x_0 +4\,{\rm Li}_2(x_0)-2 \ln^2 x_0-\frac{2\pi^2}{3}+2-\frac{12 x_0\bar x_0}{1-2x_0}\ln\frac{x_0}{\bar x_0}\right]\right\}\nonumber\\
&&-\frac{4x_0\bar x_0}{3(1-2x_0)}\frac{1}{2x_0^2}\left\{1+\frac{\alpha_s}{4\pi}C_F\left[(1+2 \ln x_0)\ln\frac{\mu^2}{M^2}-\ln x_0-14\frac{x_0}{\bar x_0}\ln x_0\right.\right.\nonumber\\
&&\left.\left.+3\frac{x_0^2}{\bar x_0^2}\ln x_0+4\, \text{Li}_2(x_0)-2\ln^2 x_0-\frac{2 \pi ^2}{3}+\frac{3}{\bar x_0}\right]\right\}\,,\\
\left\langle \frac{1}{x}\right\rangle_{3A}&=&\frac{1}{x_0}\left\{1+\frac{\alpha_s}{4\pi}C_F\left[\ln\frac{\mu^2}{M^2}\left(3+2\ln x_0\right)-(1+2x_0)\ln\bar x_0-(5-2x_0)\ln x_0\right.\right.\nonumber\\
&&~~~~~~\left.\left.+(2-8\Delta)\frac{ x_0}{\bar x_0} \ln x_0+4\,{\rm Li}_2(x_0)-2 \ln^2 x_0-\frac{2\pi^2}{3}+2\right]\right\} \,,\\
\left\langle \frac{1}{x}\right\rangle_{3A_\perp}&=&\frac{1}{x_0}\left\{1+\frac{\alpha_s}{4\pi}C_F\left[\left(3+\frac{2}{\bar x_0}\ln x_0\right)\ln\frac{\mu^2}{M^2}-(1+6x_0)\ln\bar x_0-(5-6x_0)\ln x_0\right.\right.\nonumber\\
&&\left.\left.-4\frac{ x_0}{\bar x_0} \ln x_0+\frac{1}{\bar x_0}\left(4\,{\rm Li}_2(x_0)-2\ln^2 x_0-\frac{2\pi^2}{3}\right)-\frac{8x_0\bar x_0}{1-2x_0}\ln\frac{x_0}{\bar x_0}+2\right]\right\} \nonumber\\
&&-\frac{2x_0\bar x_0}{1-2x_0}\frac{1}{2x_0^2}\left\{1+\frac{\alpha_s}{4\pi}C_F\left[\left(3-\frac{2}{\bar x_0}+\left(\frac{4}{\bar x_0}-\frac{2}{\bar x_0^2}\right) \ln x_0\right)\ln\frac{\mu^2}{M^2}\right.\right.\nonumber\\
&&~~~~~~+\left(\frac{5}{\bar x_0^2}-\frac{12}{\bar x_0}-2\bar x_0+8\right)\ln x_0+\left(\frac{4}{\bar x_0}-6+2\bar x_0\right)\ln \bar x_0\nonumber\\
&&~~~~~~\left.\left.+\frac{1-2x_0}{\bar x_0^2}\left(4\,{\rm Li}_2(x_0)-2\ln^2 x_0-\frac{2 \pi ^2}{3}\right)+\frac{1}{\bar x_0}+2\right]\right\} \,,\\
\left\langle \frac{1}{x}\right\rangle_{T}&=&-\frac{1}{2x_0^2}\left\{1+\frac{\alpha_s}{4\pi}C_F\left[(1+2 \ln x_0)\ln\frac{\mu^2}{M^2}-\left(\frac{3}{\bar x_0^2}-\frac{10}{\bar x_0}+8\right)\ln x_0\right.\right.\nonumber\\
&&\left.\left.~~~~~~+4\,\text{Li}_2(x_0)-2\ln^2 x_0-\frac{2 \pi ^2}{3}-\frac{3x_0}{\bar x_0}+3\right]\right\} \,,\\
\left\langle \frac{1}{x}\right\rangle_{T_\perp}&=&-\frac{1}{2x_0^2}\left\{1+\frac{\alpha_s}{4\pi}C_F\left[\left(3-\frac{2}{\bar x_0}+\left(\frac{4}{\bar x_0}-\frac{2}{\bar x_0^2}\right) \ln x_0\right)\ln\frac{\mu^2}{M^2}\right.\right.\nonumber\\
&&~~~~~~+\left(\frac{5}{\bar x_0^2}-\frac{4}{\bar x_0}-2\bar x_0\right)\ln x_0+\left(\frac{4}{\bar x_0}-6+2\bar x_0\right)\ln \bar x_0\nonumber\\
&&\left.\left.~~~~~~+\frac{1-2x_0}{\bar x_0^2}\left(4 \,\text{Li}_2(x_0)-2\ln^2 x_0-\frac{2 \pi ^2}{3}\right)+\frac{1}{\bar x_0}+2\right]\right\} \,,
\end{eqnarray}
\end{subequations}
with $\Delta=0$ for the NDR scheme and $\Delta=1$ for the HV scheme.
\section{Discussions and Summary\label{sect:summary}}

In this paper, we have calculated all ten twist-2 LCDAs for the S-wave and P-wave $B_c$ mesons, up to the NLO of $\alpha_s$ and the leading order of $v$. These LCDAs are re-factorized into the products of perturbatively calculable distribution parts and universal NRQCD matrix-elements for vacuum to $B_c$ state transition.  
And by use of the spin symmetry, the number of the non-perturbative NRQCD matrix-elements is reduced to two. 
Thus, such reduction of the non-perturbative inputs will potentially improve theoretical predictive power 
in hard exclusive productions of $B_c$ mesons within the framework of collinear factorization.  

On the other hand,  as we mentioned before, our results of LCDAs for $B_c$ mesons can be used to check the asymptotic behavior of the NRQCD predictions for the hard exclusive $B_c$ meson productions as in \cite{Jia:2010fw}, and resum the large $\alpha_s^n\ln^n Q^2/m_{B_c}^2$ terms by use of the ERBL equations as what done for the hard exclusive quarkonia productions in \cite{Jia:2008ep}. 

 With the excellent running of the LHC experiments, a great deal of data of $W$, $Z$ and the Higgs bosons has been accumulated. This makes it possible to study many rare and even very rare decays of $W$, $Z$ and Higgs bosons experimentally in the near future.  Recently, there are a number of theoretical researches on exclusive radiative decays of $W$, $Z$ and Higgs bosons, including decays to a photon and a quarkonium or a $B_c$ meson,  within the collinear factorization, in order to test the QCD precisely or probe the Yukawa couplings of light quarks \cite{Bodwin:2013gca,Bodwin:2014gia,Koenig:2015pha,Grossmann:2015lea,Alte:2015dpo,Bodwin:2016edd, JiaFengSang:WBc}. Thus, our results can be applied to such processes. However, we should  mention, that the study of $B_c$ exclusive production through weak charge-current is almost infeasible in experiments since the production rates is greatly 
suppressed by the CKM factor $\vert V_{cb}\vert^2\sim 10^{-3}$ \cite{JiaFengSang:WBc}
.  We think, that the study on $B_c$ exclusive productions through flavor-conserving processes, for instance $Z^0\to B_c^+ B_c^-$, is more plausible experimentally. 

Besides the exclusive processes, our results can also be applied to study the $B_c$ semi-inclusive productions in hadron colliders at relatively low $p_T$ region. Some researchers have pointed out that the double parton fragmentation mechanism may become as important as the single-parton fragmentation mechanism for quarkonia productions at colliders at relatively low $p_T$ region \cite{Kang:2011mg,Kang:2011zza,Fleming:2012wy,Kang:2014tta,Kang:2014pya}. In the corresponding factorization formula, the double-parton fragmentation functions serve as very important non-perturbative quantities. 
The color-singlet parts of these double-parton fragmentation functions for quarkonia can be related to the LCDAs of quarkonia \cite{Ma:2013yla,Ma:2014eja}.  Therefore, with the same reasoning, our results of LCDAs for $B_c$ mesons can deduce the color-singlet parts of the double-parton fragmentation functions for $B_c$ mesons.

Although we have calculated the LCDAs up to NLO of $\alpha_s$ and leading order of $v$, the corresponding relativistic corrections may be as important as, or even more important than, the radiative corrections. Thus, in order to test the applications of LCDAs in $B_c$ meson productions precisely, one should consider the relativistic corrections too. We leave this task to future work.

\section*{Acknowledgement}

The authors thank Prof. Yu Jia and Cong-Feng Qiao for enormous inspiring discussions on many issues related to this work. This work is partially supported by the National Natural Science Foundation of China under Grants No. 11275263.

\end{document}